\documentclass{aastex}
\slugcomment{Submitted to: 
The Astrophysical Journal}
\shorttitle{Gravitationally Magnified Gravity}
\shortauthors{Nemiroff}

\begin{document}

\title{Can a Gravitational Lens Magnify Gravity?  A Possible Solar System Test}

\author{Robert J. Nemiroff}
\affil{Michigan Technological University, Department of Physics, \\
1400 Townsend Drive, Houghton, MI  49931}

\begin{abstract}
Can a gravitational lens magnify gravity? Leading theories of gravity likely indicate ``no", but the possibility might be testable by using our Sun as a gravitational lens and carefully aligning a satellite past the minimum transparent focal distance of about 25 AU.  Here the magnitude of a maximal effect is estimated and found potentially observable.  
\end{abstract}


\section{Introduction}

Can a gravitational lens magnify gravity?  A naive answer based in purely Newtonian gravity would be ``no" since there, light and gravity move infinitely fast.  Slightly less naive speculation might hold the answer to be ``yes", assuming that gravity moves and acts somehow like high-frequency light.  Possibly less naive speculation holds that the answer is ``no" since gravitational distortion does not really move like light even in General Relativity (GR), and in quantum formulations of gravity, virtual particles that might carry gravity are not so restricted in speed.  

Tests of metric gravitational theories have been broken down into experimental tests of a few basic parameters \citep{Wil72}.  Expanded into parameterized post Newtonian (PPN) approximations, the two most basic parameters are $\gamma$, quantifying the amount of space curvature created by a point mass, and $\beta$, quantifying the amount of non-linearity of space curvature when point masses are superposed.  The most stringent recent test was conducted with the Cassini mission now orbiting Saturn, where the frequency shift by the Sun was measured to agree with the GR prediction of $\gamma = 1$ to an accuracy of $10^{-5}$ \citep{Ber03}. Using the Cassini result and Lunar Laser Ranging data \citep{Mul98}, $\beta$ has been found to agree with it GR prediction of $\beta = 1$ to an accuracy of $10^{-4}$ \citep{Wil04}.  Tests of the Strong Equivalence Principle (SEP) can be quantified in terms of $\eta = 4 \beta - \gamma - 3$. Using the same Solar System data also constrains variation of the gravitational constant ${\dot G}/G$ to be less than $10^{-12}$ \citep{Wil04}.  Broadly, GR has been able to explain every gravitational Solar System experiment yet \citep{Tur04}.

That $\beta$ is unity in GR, and tests near unity in Solar System measurements, is a statement that gravity itself {\it produces} gravity.  It is not clear that this has any bearing on whether gravity can {\it lens} gravity, though. There appears to be no direct calculation of this specific effect in GR or any gravitational theory at this time. As with tests of the Strong Equivalence Principle and ${\dot G}/G$, the question of whether gravity lenses gravity can be asked outside the application of any specific model for gravity. 

An odd feature of gravitationally lensed gravity is that the effect is essentially negligible inside the minimum transparent focal length of the lens, and outside a precise alignment of the source, the lens, and the observer.  This minimum transparent focal distance for the Sun is estimated to be about 24 AU \citep{Dem00}, which lies between the orbits of Uranus and Neptune.  Therefore, tests of Solar System PPN parameters even by Cassini at Saturn (9.5 AU) might have little relevance.  No other large mass in our Solar System has so short a focal length as our Sun.

It might even be possible that the falsification of such an effect could favor some formulations of gravity over others, however esoteric.  For example, if gravity actually curves spacetime then photons, particles, virtual particles, and everything else must negotiate the curvature of that spacetime. If, on the other hand, spacetime curvature is really an illusion caused by a more fundamental quantum mechanism of force on a flat background, one might expect that some interactions are immune to such spacetime curvature.  In other words, taken to an extreme, can virtual particles just completely ignore the gravity of a black hole?

For the sake of comparison, however, two extremes are assumed.  In the first, a mass gravitationally lenses gravity with the same magnification as the gravitational lens does for wavelengths of light much shorter than the Schwarzschild radius of the lens.  In the second extreme, gravity has no such gravitational lens effect on the gravitation of another body. 

In the maximal ``a gravitational lens can magnify gravity" case, regions exist where the gravity of a distant mass, the source, is magnified by a nearby mass, the lens.  Such a region would begin at the minimum focal length of the lens and extend to the horizon. This long thin tube will be called a {\it gravitational hollow} as it might appear to act like it contains a small mass when it does not. 

In this manuscript, the most prominently discussed case will involve the interior of the Sun as the gravitational lens.  In Section 2 relevant lensing formulae are estimated, and in Section 3 they are applied to our Sun.  Section 4 discusses candidate sources that our Sun might detectably magnify.  Section 5 gives a brief discussion of some possible non-Solar hollows, while in Section 6 gives a summary and some conclusions.

\section{Estimation of Relevant Lensing Parameters}

One thing that is surely known about a gravitational lens is that it can magnify light.  More specifically, that our Sun can act as a gravitational lens was confirmed by  \citet{Edd19}.  From the measured deflection value of 1.76 arcseconds at the Solar limb, it is straightforward to show that the minimum focal length of the opaque Sun is about 550 AU (see, for example, \citet{Mac94}).  High amplification regions of gravitational lensing near the lens have been discussed previously by \citet{Nem95} and \citet{Wan96}.

Lensing calculations limited to the Solar limb assume the Sun to be opaque, as it well is to optical light.  There are, however, radiations to which the interior of our Sun is not opaque.  For these, including low energy gravitational radiations and neutrinos \citep{Nem97}, the smallest focus has been estimated to be about 30 AU based on the solar model of \citet{Dzi94}.  More recently, \citet{Dem00} has estimated the minimum transparent solar focus to be about 24 AU, based on a standard solar model of \citet{Bah89}.  For the sake of estimation purposes, the minimum solar focus will be considered to be at 25 AU.  The reason that the Sun's transparent focus is so much less than its opaque focus is that the Sun is very much centrally condensed.  

In this paper, several simplifying assumptions will be made in the interest of roughly estimating the possible magnitude for these effects.  First, it will be assumed that the Sun is spherically symmetric -- no extra gravitational lensing convergence or shear will be assumed.   The center of the Sun will be assumed to act like a point gravitational lens.  All sources will be assumed to be circular and have a constant projected surface mass density.  Furthermore, it will be assumed that the wavelength of the radiations detected will be short compared to the Schwarzschild radius of the lens, so that classical estimations typically used in gravitational lens theory can be directly applied here. 

For a point source in the field of a point (Schwarzschild) lens, the flux magnification due to gravitational lensing is \citep{Ref64}
 \begin{equation} \label{A}
  A = { {u^2 + 2} \over {u \sqrt{u^2 + 4} } },
 \end{equation}
where $u$ is the angular separation between the lens and the source  as seen by the observer, in units of the angular radius of the Einstein ring (for a review of linear gravitational lensing, see, for example, \citet{Pac96}). Note that when the observer approaches perfect lens and source alignment, the magnification formally diverges. 

Were the source to have any finite size at all, the formally divergent magnification becomes finite, but may be extremely high.  More specifically, when the centers of a point lens and a uniform circular source are precisely aligned, the maximum gravitational lens amplification achieved is 
 \begin{equation} \label{Amax}
  A_{max} = [ 4 (\theta_E / \theta_*)^2 + 1 ]^{1/2} 
 \end{equation}
where $\theta_E$ is the angular size of the Einstein ring and $\theta_* = R_*/D_*$ is the angular radius of the unlensed source (see, for example, \citet{Sch92} for a good discussion).  Here $\theta_E = \sqrt{2 R_S D_{LS} / (D_{lens} D_*)}$ where $R_S$ is the Schwarzschild radius of the lens, $D_{LS}$ is the angular diameter distance between the lens and the source, $D_{lens}$ is the distance between the observer and the lens, and $D_*$ is the distance to the source star.

At high amplification of a uniform source, $\theta_E >> \theta_*$ so $A_{max} \sim 2 \theta_E / \theta_*$.  When any part of a uniform circular source is superposed directly behind the lens, the magnification is near this maximum, within about a factor of two \citep{Nem94}.  A simple way to visual this is to assume that lensing takes a uniform unlensed circle of angular area $\pi \theta_*^2$ and, under perfect gravitational lens alignment, distorts it into a lensed annulus of angular radius $\theta_E$ and angular width $\theta_*$.  The ratio of the lensed to unlensed areas is then the magnification, which is $A_{max} \sim 2 \pi \theta_E \theta_* / (\pi \theta_*^2) = 2 \theta_E / \theta_*$, as estimated above.

Now although the length of the gravitational hollow goes from the minimum Solar focus to the horizon, the width of the hollow is much smaller.  For practical purposes, the width of the hollow is estimated to be the region where any part of the source is superposed directly behind the center of the lens \citep{Nem94}.  It is then found that that
 \begin{equation} \label{width}
 w_{hollow} \sim R_* D_{lens} / D_{LS} .
 \end{equation}

It is expedient to ask what, exactly, about gravity might be magnified by a gravitational lens.  For light, lensing changes only the apparent angular area the source subtends in the observer's sky, leaving surface brightness and redshift unchanged.  One might assume that a mass $M$ magnified by an amount $A$ results in an apparent mass of $A M$.  This would be the result were virtual gravitons to follow null geodesics, ignoring the effects of redshifting.  Therefore, this is will be assumed here as the ``maximal magnification" case.

\section{Application to our Sun}

For the purposes of practical estimation for our Sun, the internal mass that produces this Einstein ring is roughly contained in about $0.2 R_{\odot}$ \citep{Dem00}, which is where the Einstein ring would appear.  Maximum magnification by the transparent Sun would then be 
 \begin{equation} \label{Amax2}
 A_{max} \sim 2 \left( { 0.2 R_{\odot} \over 25 AU } \right)
              \left( {D_* \over R_*}               \right)
         \sim 2 \left( { 0.2 R_{\odot} \over R_*} \right) 
              \left( { D_* \over 25 AU }            \right)
         \sim  2 \ {\rm x} \ 10^4 \ 
              \left( {0.2 R_{\odot} \over R_*} \right) 
              \left( {D_* \over 1 \ {\rm pc} } \right)  .
 \end{equation}

When further than the minimum focal distance of the transparent Sun, near maximum magnification will be recorded for observers for which the center of the lens is superposed on the unlensed source \citep{Nem94}.  The width of resulting gravitational hollow is therefore estimated to be
 \begin{equation} \label{width2}
 w_{hollow} = 85 \ {\rm km} \ 
              \left( { R_*  \over R_{\odot} } \right)
              \left( {1 \ \rm{pc} \over D_*  }  \right) .
 \end{equation}

\subsection{Inside the Hollow: Anomalous Gravitational Acceleration}

Is this magnification measurable?  Suppose now that a spacecraft enters the gravitational hollow of the Sun opposite a background star.  Described classically, the acceleration caused by the lens toward the lens would be $a_{lens} = G M_{\odot} / D_{\odot}^2$.  Without the gravitational hollow, the acceleration caused by the distant star toward the distant star (nearly the lens center, too) would be $a_* = G M_* / D_*^2$.  Inside the gravitational hollow, a maximally magnified acceleration of the spacecraft toward the distant star would be $a_{hollow} \sim G (A_{max} M_*) / D_*^2$.  In this case $\Delta a / a = [a_{hollow} + a_{lens}) - (a_* + a_{lens})] / (a_* + a_{lens}) \sim A_{max} a_* / a_{lens}$ so that
 \begin{equation} \label{da1}
  {\Delta a \over a } \sim A_{max} 
                           \left( {M_* \over M_{lens} } \right)
                           \left( {D_{lens} \over D_* } \right)^2 .
 \end{equation}
For the minimum focal length of the transparent Sun, it is then seen that 
 \begin{equation} \label{da2}
 {\Delta a \over a} \sim 10^{-8} \
                         A_{max} \ 
                         \left( {M_* \over M_{\odot} } \right) 
                         \left( {1 \ \rm{pc} \over D_* } \right)^2 .
 \end{equation}
Substituting in for $A_{max}$ then gives
 \begin{equation} \label{da3}
 {\Delta a \over a} \sim 2 \ {\rm x} \ 10^{-4} \ 
                         \left( {0.2 R_{\odot} \over R_*} \right) 
                         \left( {M_* \over M_{\odot} } \right)
                         \left( {1 \ \rm{pc} \over D_* } \right).
 \end{equation}

Were a spacecraft to sit in the hollow, it might experience this small addition to its sunward acceleration.  Were the spacecraft able to sit in the hollow for a sufficiently long period, such an acceleration might be detectable.

\subsection{Inside the Hollow: Anomalous Redshift and Time Dilation}

Alternatively, a spacecraft that sits in the Sun's gravitation hollow might be expected to show a gravitational redshift for radiation emitted out of the hollow, in particular toward Earth.  Only the relative redshift between a spacecraft inside the hollow and Earth, and one just outside and Earth, will be considered.  The expected difference in wavelength due to the gravitational redshift would be $\Delta \lambda / \lambda \sim R_{S*} / 2 D_*$ where $R_{S*} = 2 G M_* / c^2$ is the Schwarzschild radius of the distant star, $G$ is the gravitational constant, and $c$ is the speed of light.  The observer at the Earth is always assumed to be at the same distance from the spacecraft --  effectively at an infinite distance from the lensed star.

A practical way to think about the effective gravitational redshift from the lensed star in the hollow is to consider that its gravitational redshift would be the same as if the spacecraft was only $D_E* = D_*/A_{max}^{1/2}$ away from the lensed star.  One would then expect that
 \begin{equation} \label{dl1}
 {\Delta \lambda \over \lambda}  \sim  {A_{max}^{1/2} R_{S*} \over D_* }.
 \end{equation}

Putting this in a more convenient form,
 \begin{equation} \label{dl2}
 {\Delta \lambda \over \lambda} \sim 10^{-16} \ 
                                     A_{max}^{1/2} \
                                     \left( {M \over M_{\odot}} \right)
                                     \left( {1 \ \rm{pc} \over D_* } \right) .
 \end{equation}
Including the above formula for magnification gives 
 \begin{equation} \label{dl3}
 {\Delta \lambda \over \lambda} \sim 2 \ {\rm x} \ 10^{-14} \
                  \left( {M \over M_{\odot} } \right)
                  \left( {0.2 R_{\odot} \over R_*} \right)^{1/2} 
                  \left( {1 \ \rm{pc} \over D_*} \right)^{1/2} .
 \end{equation}

A spacecraft in the hollow might expect to have its radiation redshifted by this amount.  Similarly, the amount of time dilation, $\Delta t / t$ for pulses emitted by the spacecraft in the hollow, relative to pulses emitted for pulses just outside the hollow, is of the same order.

\subsection{Across the Hollow: Anomalous Time Delay}

Were very precise signals sent across the hollow, slight deviations in timing created by the hollow might be detectable.  This might a relatively easy way to test for the existence of a hollow, since positioning a satellite in a hollow might be relatively difficult.  A spacecraft could be instructed to drift behind a range of possible positions of the hollow with constant watch on for a slight deviation in the arrival time of a time signal.  Such deviations are analogous to the Shapiro time delay measured for satellites across our Solar System \citep{Sha77}.

Note that to a first approximation, a comparison of pulses emitted that miss the hollow to those that go through the hollow should show no difference in redshift or the rate of arrival times.  This is because whatever redshift occurred for pulses coming out of the hollow, an equivalent blueshift should exist for pulses entering the hollow.  A hollow would rather create a phase shift, or ``temporal drift" in the arrival times of pulses.

The amount of temporal drift is estimated as follows.  Assume that the relative rate the time runs slow in the hollow is $\Delta \lambda / \lambda$ as in equation (\ref{dl1}), since $\lambda \propto t$ for light.  Assume this rate is constant across the hollow, so that $\delta t / t \sim$ constant.  The time shift should then be the integral of the time lost by the photons as they cross the hollow.  This means it should be proportional to the time it takes light to cross the hollow.  Therefore, 
 \begin{equation} \label{tdrift1}
 t_{drift} = \int_0^{t_{cross}} (\Delta t / t) \ dt = t_{cross} (\Delta t / t) .
 \end{equation}
Now $t_{cross}$ is just the time it takes for a pulse to cross a hollow.  Assuming it goes straight across, and assuming for estimation purposes that the rate of time slowing is constant in the hollow, then $t_{cross}$ is just width of the hollow divided by the speed of light, assuming a minimal crossing.  Therefore $t_{cross} \sim R_* D_{\odot} / (c D_*)$ so that 
 \begin{equation} \label{tdrift2}
 t_{drift} \sim { A_{max}^{1/2} R_* D_{\odot} R_{S*} \over 
                 c D_*^2 }  .
 \end{equation}
so that
 \begin{equation} \label{tdrift3}
 t_{drift} \sim 2.7 \ \rm{x} \ 10^{-17} \ {\rm sec} \ 
                A_{max}^{1/2} 
                \left( {R_* \over R_{\odot} } \right) 
                \left( {M_* \over M_{\odot} } \right)
                \left( {1 \ \rm{pc} \over D_*} \right)^2 .
 \end{equation}
Substituting the estimation for $A_{max}$ above gives
 \begin{equation} \label{tdrift4}
 t_{drift} \sim 2.4 \ \rm{x} \ 10^{-15} \ {\rm sec} \
                \left( {R_* \over R_{\odot} } \right)^{1/2}
                \left( {M_* \over M_{\odot} } \right) 
                \left( {1 \ \rm{pc} \over D_* } \right)^{3/2} .
 \end{equation}
It is evident that $t_{drift}$ is actually quite sensitive to the width of the hollow and so the distance to the source $D_*$.

\section{Candidate Sources}

Several candidate sources of mass are now considered that might have their gravity detectably lensed by our Sun.  These include the brightest star, the closest white dwarf, the closest neutron star, and the supermassive black hole thought to reside at our Galactic Center.  Here are their parameters:

\begin{deluxetable}{llrrr}
\tabletypesize{\scriptsize}
\rotate
\tablecaption{Attributes of candidate source stars}
\tablewidth{0pt}
\tablehead{
\colhead{Name} & \colhead{Type} & \colhead{$D_*$ (pc)} & \colhead{$R_*/R_{\odot}$} & \colhead{$M_*/M_{\odot}$} 
}
\startdata
Sirius A        & Brightest Star       & 2.7   & 1.7 &     2.1 \\
Sirius B        & Closest White Dwarf  & 2.7   & 8.4E-03 & 1.0 \\
J185635-3754    & Closest Neutron Star & 61.   & 1.4E-05 & 1.4 \\
Galactic Center & Black Hole           & 8100. & 14.     & 3.0E+06 \\
\enddata
\end{deluxetable}

From these cases, attributes of the possible gravitational hollows are roughly estimated. Potentially measurable variables are estimated only for spacecraft at or near the innermost focal point of the transparent Sun.  

\begin{deluxetable}{lrrrrr}
\tabletypesize{\scriptsize}
\rotate
\tablecaption{Attributes of resulting gravitational hollows}
\tablewidth{0pt}
\tablehead{
\colhead{Type} & \colhead{$A_{max}$} & \colhead{$w_{hollow}$ (km)} & 
\colhead{$\Delta a / a$} & \colhead{$\Delta \lambda / \lambda$} & \colhead{$t_{drift}$ (sec)} 
}
\startdata
Brightest Star &  6.4E+03 &  54.     &  9.2E-06 &  2.0E-15  &  7.4E-16 \\
Closest WD     &  1.3E+06 &  0.26    &  8.8E-04 &  1.3E-14  &  2.5E-17 \\
Closest NS     &  1.7E+10 &  2.0E-05 &  3.2E-02 &  9.4E-14  &  1.3E-20 \\
GC Black Hole  &  2.4E+07 &  0.15    &  5.3E-04 &  1.8E-11  &  1.8E-14  \\
\enddata
\end{deluxetable}

Inspection of the results in Table 2 is interesting.  The highest magnifications are evident for the nearest neutron star, and hence, by implication, for nearby neutron stars and black holes of stellar origin.

Is any of this measurable?  As far as gravitational redshift goes, $z = \Delta f / f = \Delta l / l$ can be measured only to an accuracy that matches a clock stability.  Now the outer Solar System Voyager probes used an ultra-stable oscillator with a frequency stability of about $10^{-12}$ \citep{Kri94}.  However, \citet{Kri94} has claimed that using a trapped ion frequency standard, an outer Solar System test could measure $z$ to an accuracy of $z \sim 10^{-16}$ for a time period of weeks.  If such a gravitational redshift could be recorded to that accuracy by a spacecraft in the hollow and Earth, any of the gravitational hollows listed in Table 2 above would be detectable.  

Since $\Delta l / l = \Delta t / t$, the $\Delta l / l$ column in Table 2 can also be used to estimate whether the time dilation on a clock placed in a gravitational hollow is measurable.  Currently the most precise clock on Earth is the NIST-F1 clock which keeps time with a precision of one part in $10^{15}$
\citep{Jef01}.  Therefore all of the gravitational hollows estimated above should be detectable.  In particular, the hollow opposite the supermassive black hole in the Galactic center should be several orders of magnitude above detectability.

Next, finding small changes in acceleration might well be doable but the detectability likely depends on how long an experiment can sit in a gravitational hollow.  As an example detection scenario, timing pulses from a satellite in the hollow would be expected to be offset from their expected times as the spacecraft becomes slightly displaced by the anomalous acceleration.  More specifically, the offset distance would be $\Delta d = (1/2) \Delta a t^2$ so offset time would be $t_{offset} = \Delta d / c = (\Delta a /a) a t^2 / (2 c)$.  Therefore, assuming that $a$ and $\Delta a$ are approximately constant over the spacecraft trajectory, then
 \begin{equation} \label{toffset1}
 t_{offset} \sim  { (\Delta a / a) G M_{lens} t^2 \over
                   2 c D_{lens}^2 }  .
 \end{equation}
Near the innermost part of the Sun's gravitational hollow 
 \begin{equation}  \label{toffset2}
 t_{offset} = 1.6 \ {\rm x} \ 10^{-14} \ 
                  \left( {\Delta a \over a } \right)
                  \left( {t \over 1 \ {\rm sec} } \right) .
 \end{equation}
Given a spacecraft drifting in the hollow for one week, all of the hollows in Table 2 would be detectable.  Practically, however, the NS hollow might be too small to place a sensitive experiment.

Lastly, the time drift for a distant clock behind the halo might also be detectable.  The problem with practical measurement is keeping the source behind the hollow for a long enough time to measure $t_{drift}$.  A background source like a pulsar might have the needed time accuracy over many pulses but cross behind the hollow in too short a time to actually measure the effect.

Even if hollows exist and are detectable in principle, they might not be locatable.  To find the position of a hollow, one must know the angular position of its source on the sky to a precision that allows a spacecraft to maneuvering into the hollow.  More precisely, the angular precision needed would be on order $\alpha \sim h_{width}/D_{lens}$.  For the case of the hollow beginning at the minimum focal distance of the transparent Sun, the lens is assumed to be at about 25 AU.  Therefore, for the cases of the Sirius A, Sirius B, the closest neutron star, and the Galactic central black hole, $\alpha =$ 2.9 x $10^{-3}$, 1.5 x $10^{-5}$, 1.1 x $10^{-9}$, and 8.1 x $10^{-6}$ arcseconds respectively.  Now the Hipparcos satellite can measure angular positions to an accuracy of $10^{-3}$ arcseconds \citep{Esa97}, and the planned Space Inteferometry Mission (SIM) for 2009 is expected to measure angular distances to an accuracy of about 4 x $10^{-6}$ arcseconds.  Therefore, all but the neutron star hollow should be locatable.  Were a spacecraft or distant pulsar to drift behind a hollow, the need for high positional accuracy would be alleviated.

\section{Searching for Distant Hollows}

Most of the above discussion has focused on detecting the gravitational hollows created by our own Sun.  All massive objects create hollows, though, and a clever observing scheme might uncover one out in the more distant universe. 

The possible scenarios are too numerous to mention, and surely not every scenario has been thought of, but two of the more obvious ones will be mentioned here.  One interesting case is the set of hollows that would surround a black hole. These hollows would extend right up to the event horizon. In the case where one black hole orbits another, a hollow of magnified gravity will follow each black hole around.  One should note that when even in the maximum case where one black hole gravitationally lenses the gravity of another black hole, the resulting hollow of enhanced gravity does not create infinite magnifications nor induce infinite accelerations.  Rather, in a hollow, the mass of the more distant black hole will appear to be increase by a factor of roughly $A_{max} \sim 2 \theta_E / \theta_*$.  

Stars near a large black hole such as expected near the centers of galaxies would likely create large gravitational hollows of increased gravitational curvature in the directions opposite the supermassive central black hole.  Gas, stars, planets, and even asteroids might expect regions of suddenly high tidal effects when passing through such regions.  In some extreme cases, objects might be disrupted. 

Another case is to consider all of the gravitational hollows moving about in the universe collectively.  Rather than living in a relatively calm gravitational sea, any object passes in or near a multitude of gravitational hollows continually.  Some distant star might have its gravity magnified by another distant star by a large factor for a brief period, which by itself might be undetectable.  Possibly, however, a ``gravitational jitter" of all of these combined would be detectable by a sufficiently accurate clock.

\section{Summary and Discussion}

Were a gravitational lens to magnify gravity, peculiar effects might exist.  One might wonder weather such an effect would allow for the production of a perpetual motion machine.  One example might be a wheel with paddles where one paddle is inside the gravitational hollow and feels a relatively high acceleration toward the aligned masses, while another paddle is outside the hollow and feels a relatively low acceleration, resulting in a net spin of the wheel.  Although such conjecture is troubling, it might not disqualify the effect inherently.

For another possible peculiar effect, picture two masses approaching a black hole from opposite directions.  As they together cross the black hole's photon sphere, the {\it direction} of the gravity felt by each mass from the other mass would be in the opposite direction from the black hole.  In the presence of the black hole, the masses would otherwise gravitationally repel!  Given that a black hole might be considered as mass hovering just outside the event horizon, it might be that if gravity lenses gravity, the Schwarzschild metric itself is not an accurate description of gravity near the event horizon.

For the transparent Sun, it is possible that a region of magnification larger than $A_{max}$ of equation \ref{Amax} exists.  The magnifications discussed above were all derived assuming a point lens that effectively constrained the angular width of the Einstein ring to be that of the unlensed angular radius of the source.  This is certainly true for geodesics that cross far from the Sun's center and focus outside the minimum solar focus. However, the density profile near the center of the Sun might create regions where a wider range of null geodesics are focussed to a single distance, creating an apparent Einstein ring width much greater than the point lens case. Such rings could show magnifications as high as $A_{max} \sim (\theta_E / \theta_*)^2$.

Note that a gravitational hollow should not be considered a created region of new gravity, but rather an angular redistribution of already existing gravity.  In other words, the excess gravity felt by a mass inside a gravitational hollow is made up for by a very slightly decreased gravity felt by all other masses outside the hollow.

Last, it is noted that four spacecraft are already outside 25 AU, the minimum focal length of the transparent Sun.  These are the two Pioneer and two Voyager spacecraft launched by NASA last century.  The existence of these spacecraft may be taken as demonstrations that it is already technologically possible to reach the minimum focal distance of the transparent Sun.

\acknowledgements
Bijunath Patla, Christ Ftaclas and Amos Ori are thanked for useful conversations.

\end{document}